## Research Article
# Anisotropic Diffusion for Details Enhancement in Multiexposure Image Fusion

### Harbinder Singh,[1] Vinay Kumar,[2] and Sunil Bhooshan[3]

[1] *Baddi University of Emerging Sciences and Technology, Baddi, Solan 173205, India*
[2] *Grupo de Procesado Multimedia, Departamento de Teoría de la Señal y Comunicaciones, Universidad Carlos III de Madrid, Leganés, 28911 Madrid, Spain*
[3] *Jaypee University of Information Technology, Waknaghat, Solan 173215, India*

Correspondence should be addressed to Harbinder Singh; harbinder.ece@gmail.com

We develop a multiexposure image fusion method based on texture features, which exploits the edge preserving and intraregion smoothing property of nonlinear diffusion filters based on partial differential equations (PDE). With the captured multi-exposure image series, we first decompose images into base layers and detail layers to extract sharp details and fine details, respectively. The magnitude of the gradient of the image intensity is utilized to encourage smoothness at homogeneous regions in preference to inhomogeneous regions. Then, we have considered texture features of the base layer to generate a mask (i.e., decision mask) that guides the fusion of base layers in multiresolution fashion. Finally, well-exposed fused image is obtained that combines fused base layer and the detail layers at each scale across all the input exposures. Proposed algorithm skipping complex High Dynamic Range Image (HDRI) generation and tone mapping steps to produce detail preserving image for display on standard dynamic range display devices. Moreover, our technique is effective for blending flash/no-flash image pair and multifocus images, that is, images focused on different targets.

## 1. Introduction

It is impossible to capture the entire dynamic range of the real world scene with single exposure. Human eye is sensitive to relative rather than absolute luminance values [1]. Human eye can observe both indoor and outdoor details simultaneously. This is because the eye adapts locally as we scan the different regions of the scene and can adapt 10 orders of magnitude of intensity variations in the scene [2], while standard digital cameras are unable to record the luminance variation in the entire scene. Currently, there are many applications that involve variable exposure photography to determine the details to be captured optimally in the photographed scene. The intention of exposure setting determination is to control charge capacity of the Charge Coupled Device (CCD). An example is shown in Figure 1(a), and long exposure yields details in the poorly illuminated areas while short exposure provides detail in the brightly illuminated area. Therefore, each exposure gives us trustworthy information about certain pixels, that is, the optimally exposed pixels for that image. In such type of images, for dark pixels, the relative contribution of noise is high and for bright pixels, the sensor may have been saturated. Therefore, it is desirable to ignore very dark and very bright pixels to achieve suprathreshold viewing conditions [2]. Consequently, the scene contains very dark and very bright areas which are partially under- or overexposed in the optimally exposed photograph (see Figure 1(a)). This is because of limited dynamic range (DR) of the standard digital cameras (i.e., $10^2$). The solution is to photograph the scene several times with variable exposures and reconstruct blended image that contains the whole details, even in brightly and poorly illuminated areas. High dynamic range imaging (HDRI) [3–7] techniques give the solution to recover radiance maps from photographs taken with conventional imaging equipment.

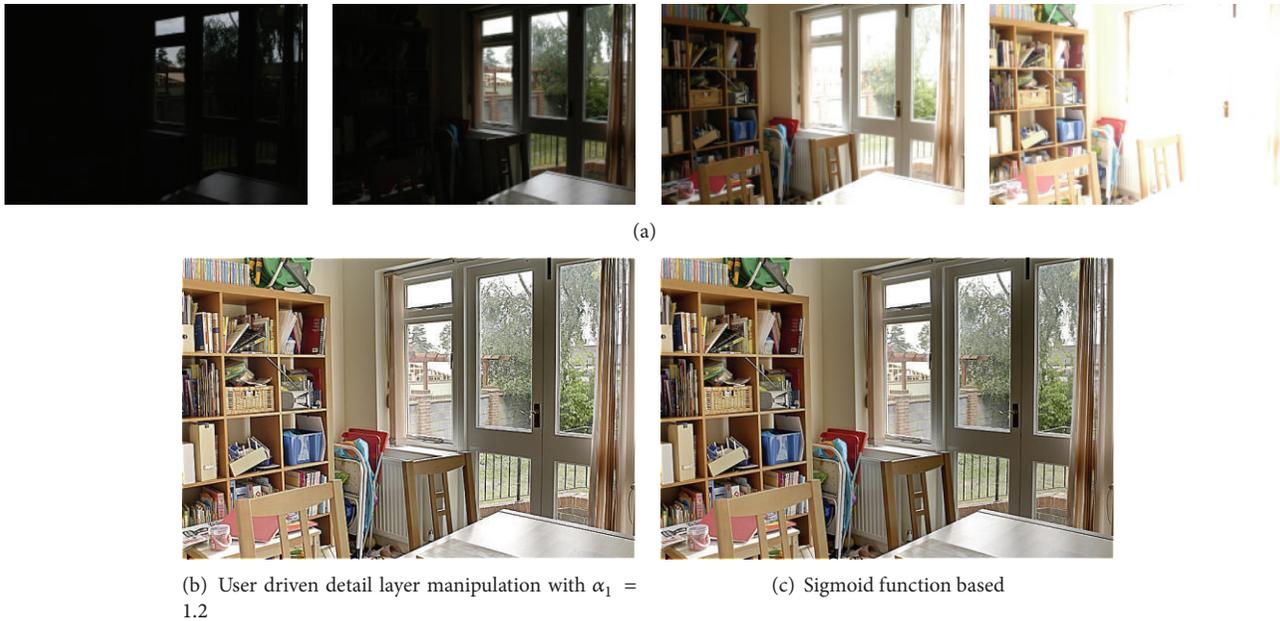

(a)

(b) User driven detail layer manipulation with $\alpha_1 = 1.2$

(c) Sigmoid function based

Figure 1: (a) Images representing multiple exposures; ((b), (c)) illustration of our detail-preserving exposure fusion result. Note that our result contains more details in brightly and poorly illuminated areas with natural contrast. The fine textures on the chair are accurately preserved. (c) Our detail layer enhancement based on sigmoid function across all the inputs reveals more texture details in the fused image and does not depict noticeable artifacts near strong edges.

To make the concept of dynamic range clear, let us redefine some useful terms. Image is said to be low dynamic range (LDR) when its dynamic range is lower than that of the output medium. A standard dynamic range (SDR) image is the one whose dynamic range corresponds approximately to that of the standard output medium (i.e., 0–255 OR about $10^2$) and is called display-referred image. A high dynamic range (HDR) image has dynamic range higher than that of the output medium and it is called scene-referred image. Alternatively, the standard displays (LCD, CRT) and printers have limited contrast ratio (i.e., dynamic range). Therefore, these devices are unable to reproduce full dynamic range that leads to tone mapping problem. Tone mapping [8] is the technique to remap the intensities for display HDR images on SDR devices. Although few HDR display devices have been developed and will become generally available in the near future, this technology is very expensive and not accessible by the most users. To display HDR data directly, a number of HDR display prototypes are proposed recently by [1, 9, 10]. As a result, there will always be a need to prepare HDR imagery for display on LDR devices or directly generate an image that looks like tone-mapped image [1]. Consequently, we need efficient exposure fusion technique to preserve scene details without intermediate representation. The goal of exposure fusion mechanism is to maximize information content of the synthesized scene from a set of multiexposure images without computing HDR radiance map and tone mapping (see Figures 1(b) and 1(c)).

Compositing is done on the pixel intensity values rather than irradiance values. This approach does not care about the exposure times and camera response function (CRF), which is required to linearize the image data before combining LDR exposures into HDR image [3]. Following the consideration of pixel intensity based fusion, the major focus of this paper is the utilization of conceptually simple, computationally simple, and robust texture features, specifically local range of base layer, for the identification of well-exposed regions. The base layers across all input images are fused by using multiresolution pyramid approach [11] to preserve local spatial structure that provides high quality spectral content in the fused image. We have considered texture features of the image to generate a mask that guides the fusion of base layers computed across all the input images. The base layer is computed by applying nonlinear filter [12] that preserves the locations where the magnitude of the gradient has maximum value and the detail layer is then computed as the difference between the original input image and the base layer. The algorithm overcomes the major drawbacks of conventional multiresolution pyramid based fusion [13], namely, the blurring of edge details and the introduction of artifacts.

A first step, in our algorithm, is multiscale decomposition (MSD) of each image to extract details at arbitrary scales, based on adaptive and edge preserving filter (i.e., anisotropic diffusion) [12]. Our algorithm takes $N$ identically sized multi-exposure images taken from a fixed viewpoint and produces output image of the same size, in which well-exposed pixel value is computed by combining detail information from all of the input images at each scale of the decomposition. Unlike earlier image-based compositing techniques [13], our approach separates coarse scale details (i.e., base layer) from fine details (i.e., detail layer), while our approach is similar in spirit to the multiscale shape and detail enhancement from multilight image collections (MLIC) approach of Fattal et al. [14]. Therefore, our approach is effective to control fine and coarse details separately during the compositing process and needs no further postprocessing. After the manipulation of each redundant layer and fused base layer, the detail layers across all input images are recombined to produce well-exposed image (see Figure 2). Thus, the magnitude of the base layer is modified based on the decision map to ensure that resulting fused image contains well-exposed regions, while the magnitude of the detail layer is unchanged, thereby preserving detail. To be able to deal with strong edges separately, we use a nonlinear multiscale edge-preserving image decomposition which permits us to manipulate and combine details at multiple scales without introducing visible halos and artifacts.

Although the proposed framework does not require human intervention, in practice, we provide set of parameters in Section 4 that allow users to interactively control the detail enhancement in the fused image. The rest of this paper is organized as follows. A comprehensive review of previous work related to exposure fusion and HDR generation is provided in Section 2. Section 3 presents a description of two-scale decomposition based on ASD, texture features (i.e., local range) of the base layer that provides the weight map to guide the fusion process, and the multiresolution decomposition that reconstruct a single well-exposed base layer from a set of given multiple exposures acquired from the static scene. Section 4 illustrates the experimental results and the comparison with the popular exposure fusion and tone-mapping operators. Section 5 discusses future directions for this work and concludes this paper.

## 2. Previous Works

Image fusion techniques blend information present in different images into a single image. Burt and Adelson [11] first introduced the idea of image fusion based on Laplacian pyramid. Image fusion techniques are generally classified into three categories: pixel level, feature level, and decision level, which are reviewed by Smith and Heather [15]. Standard capturing devices can only capture either detail present in the poorly illuminated or brightly illuminated regions. Debevec and Malik [3] and Mann and Picard [4] proposed a HDRI to record the entire range of the scene radiances from different exposures that were acquired with a standard camera. Various possible formats to store radiance maps are described by Reinhard et al. [1]. "Floating point tiff" can encode a very high dynamic range (~79 orders of magnitude) without losing information.

Unfortunately, HDR images cannot be displayed on ordinary display devices with limited dynamic range. Many different global operators [1, 16–18] and local operators [8, 19–21] have been suggested for dynamic range reduction for displaying HDR images on standard display devices. Global operators apply spatially uniform remapping function on every pixel independently. For the local operators, different operations based on adaptation of human visual system are applied to different pixels. However, global operators are computationally simple than local operators. Most of the tone mapping algorithms suffer from halo artifacts and require human intervention in the parameter adjustment process. Transform domain tone mapping approaches [22] became popular compared to intensity domain. Dynamic range compression based on the properties of human visual system in gradient domain [22] is almost free of halo artifacts and require no manual parameter tweaking. They involve the gradient manipulation of local neighboring pixel at various scales to simulate adaptation behavior of human visual system. Then the image is reconstructed by solving the Poisson equation on the modified gradient fields. Recently frequency based algorithm [21] typically decomposes HDR image into base layer and detail layer. Only the magnitude of the base layer is compressed in the log domain, thereby preserving detail. The base layer of input HDR image is computed using an edge-preserving filter called the bilateral filter and the detail layer is the division of the input intensity by the base layer. The detailed review of various tone-mapping operators is given by Reinhard et al. [1].

In recent years, various fusion algorithms have been developed to assemble information from several source images to extend the depth-of-field and dynamic range of the fused image. However, the large variations in the source images, such as exposure value, focusing, modality, and environmental conditions, often make fusion extremely challenging. Ogden et al. [23] has proposed the use of Gaussian and Laplacian pyramid for image fusion. The Laplacian pyramid representation expresses an image as a sum of spatially band-passed images while retaining local spatial information in each band [11]. Image Gradient based fusion [24] provides the solution to handle strong highlights and remove self-reflections from flash and ambient images [25]. Li and Yang [26] described region segmentation and spatial frequency based multifocus image fusion. Weighted nonnegative matrix factorization and focal point analysis based multifocus fusion method [27] has been proposed to preserve feature information in fused image.

Raman and Chaudhuri [28] have utilized edge-preserving filter (i.e., bilateral filter) for the fusion of multiexposure images, in which appropriate matting function is generated based on local contrast for automatic compositing process. Image entropy based exposure fusion method was proposed by Goshtasby [29], in which an image is considered best-exposed within an area if it carries more information about the area than any other image. The optimal block size and width of the blending functions were determined using a gradient-ascent algorithm to maximize information content in the fused image. The optimal block size was varied from image to image. Images representing scenes with highly varying reflectances, highly varying surface orientations, and highly varying environmental factors such as shadows and specularities produce smaller optimal block size.

Unlike previous multiexposure fusion method proposed by Goshtasby [29], our approach calculates local range within a fixed 3-by-3 block size that reduces complexity for computing weight function to control the contribution of pixels from input bracketed images. Szeliski [30] produces fused image with improved uniformity in exposure and tone based on simple averaging the pixel brightness levels across autobracketed shots. Multi-dimensional histogram was used to analyze a set of bracketed images that projects pixels onto a curve that fits the data. Histogram equalization was used as postprocessing operator for optimal contrast enhancement in the fused image.

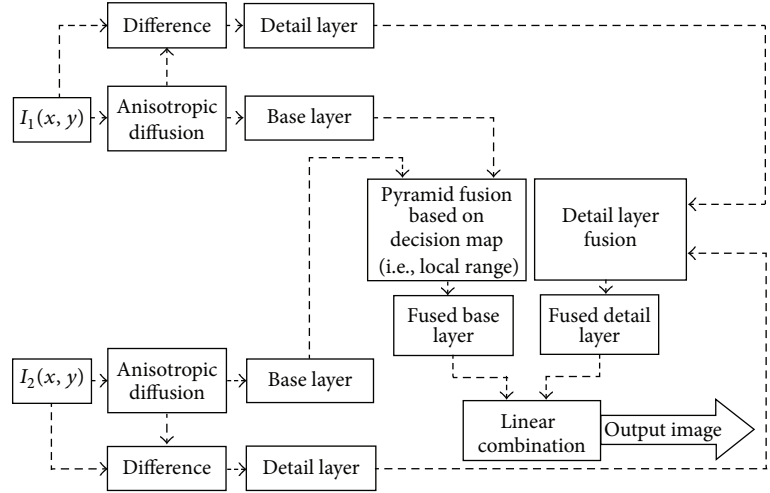

Figure 2: Proposed image domain fusion framework. Observation model illustrating the conceptual framework of the proposed texture feature based pyramid fusion approach. Note that for the concept simplicity here we have generated base layers and detail layers of two input exposures.

Recently Mertens et al. [13] propose a technique for fusing a bracketed exposure sequence into a high quality image, without converting to HDR first, which is processed based on Laplacian Pyramid. In that technique, "good" pixels are selected from image sequence guided by simple quality measures such as saturation, well-exposedness, and contrast. Zhao et al. [31] introduced a Quadrature Mirror Filters (QMFs) based subband approach for exposure fusion. Modified subbands based on calculated gain control maps according to image appearance measurements such as exposure, contrast, and saturation are blended to remove nonlinear distortion.

A number of non-adaptive MSD techniques have been proposed recently [32–34] and have some limitations. The first one is the introduction of distortions including halos and visible artifacts. Secondly, it fails to preserve edges during the decomposition. The effectiveness of edge preserving image coarsening has been recognized as valuable tool for MSD decomposition. Recently, the edge preserving MSD in [35–40] has been widely used by the graphics researchers for the image processing and the computational photography applications. Weighted least square (WLS) [36], bilateral filter (BLF) [41], anisotropic diffusion (ASD) [35], and guided image filter [40] are the popular MSD computation techniques. Among these, BLF and ASD are the well-posed approaches for preserving edges while the textures are smoothed out. BLF was first proposed by Tomasi and Manduchi [42] in 1998. The BLF is an adaptive smoothing framework that does a weighted sum of the pixels in a local neighborhood; the weights depend on both the spatial domain and the intensity domains which are used to manipulate smooth regions while preserving strong edges. Bilateral filter based exposure fusion introduced by Raman and Chaudhuri [28] uses the concept of local contrast [42] to preserve edge details. The edge-preserving MSD proposed by Perona and Malik [12] advocates the utilization of heat conduction PDE: $\partial I(x, y, t)/\partial t = \text{div}(\nabla I)$. That is, the intensity $I$ of each pixel is seen as heat and is propagated over time to its 4 neighbors according to the heat spatial variation.

In this paper, we exploit anisotropic diffusion for the fusion of images captured at different exposure settings. The base layer and detail layers are fused separately to preserve texture details. In Section 3, we will discuss the two-scale decomposition of input exposures for base layer and detail layer extraction in detail. Our technique is flexible enough to fuse flash/no-flash images and images focused on different targets (multifocus images), whereas methods proposed in [24, 25] and [26, 27] are specifically designed for the fusion of flash/no-flash and multifocus image series, respectively.

## 3. Proposed Algorithm

The objective of our exposure fusion approach is to preserve details in both brightly and poorly illuminated areas that significantly improve the quality of the fused image. It must provide optimal contrast within the capabilities of the conventional displaying medium and must not lead to artifacts such as contrast reversal or black halos. Additionally, it should produce realistic and pleasant images. The principal characteristic of our exposure fusion is an adaptive adjustment of local spatial information in the Laplacian pyramid [11] depending on texture features (i.e., local range). To control the contribution of pixels, we calculate weight that depends on the maximum and minimum intensities of the neighboring pixels from the pixel under consideration. The weight function and Gaussian-Laplacian pyramid are derived in the following sections. Figure 2 shows that the proposed scheme contains three steps, which are *analysis, scene detail manipulation based on decision map,* and *synthesis.*

More specifically, the goal of our exposure fusion algorithm to produce well-exposed image by combining the information across all of the input multiexposure images. In our implementation, two-scale decomposition based on anisotropic diffusion [12] is used to separate coarser and finer details from each input image. The base layer ($B_k$) and detail layer ($D_K$) across all $N$ input images are defined as

$$B_k = \text{aniso}(I_k), \qquad D_k = I_k - B_k, \qquad (1)$$
$$\text{where } k = 1, \ldots, N.$$

The well-exposed image is generated as

$$I_f = B_f + D_f, \qquad (2)$$

where $B_f$ is the fused base layer that maximizes the coarser details across all of the input base images $B_1, B_2, \ldots, B_N$ and $D_f$ is the residual (i.e., fused detail layer) that maximizes the finer details across all of the input detail layers $D_1, D_2, \ldots, D_N$. Before introducing the proposed approach, we briefly introduce anisotropic diffusion used to create two-scale decomposition and local range used to generate weight map for nonuniform scaling to control contribution of pixels from base layers across all of the input exposure.

### 3.1. Data Acquisition and Two-Layer Decomposition

*3.1.1. Scene Data Acquisition.* Conventional digital photography struggles with the high contrast scenes and can capture brightest part (i.e., highlights) by choosing a low exposure level (i.e., short exposure time) or the darkest part (i.e., shadows) by choosing a high exposure level (i.e., long exposure time). The information present in the fused LDR output depends on the number of input exposures captured at different exposure settings. We assume that all input multiple exposure images are photographed from static scene with the help of tripod to avoid any spatial and global misalignment. To apply our technique successfully, sequence of exposures is captured from a scene with very dark and very bright details. The aperture priority and the camera's white balance are fixed for the entire sequence. Sample input set of images with different exposure settings is illustrated in Figure 1(a).

*3.1.2. Edge Preserving Anisotropic Diffusion.* Anisotropic diffusion has led to an efficient new field to remove noise from an image by modifying the image via a Partial Differential Equation (PDE). The goal of edge preserving diffusion [12] is to encourage smoothing at homogeneous region in preference to inhomogeneous region (i.e., edge). Mathematically, the isotropic diffusion equation $\partial I(x, y, t) = \text{div}(\nabla I)$ is replaced with

$$\frac{\partial I(x, y, t)}{\partial t} = \text{div}\left[g(\|\nabla I\|)\nabla I\right], \qquad (3)$$

where $\nabla I$ is the image gradient, $\|\nabla I\|$ is the magnitude of the gradient of image intensity, $g(\|\nabla I\|)$ is an "edge-stopping function" or "conduction coefficient" that controls the diffusion strength, $(x, y)$ specifies spatial position, and $t$ is the process ordering time parameter.

The diffusion strength in the image is influenced by the conduction coefficient which depends on the magnitude of the gradient of the image intensity. The process of gradient computation from the neighbors in 1D and 2D structure is illustrated in Figures 3(a) and 3(b), respectively. If the conduction coefficient is replaced by a constant value (i.e., $g(\cdot) = 1$), the diffusion process will be isotropic linear diffusion that leads to Gaussian smoothing. Since isotropic diffusion does not consider image structure, fine textures as well as edges are smoothed. Thus for anisotropic diffusion the conduction coefficient is chosen to satisfy $g(x) \rightarrow 0$ when $x \rightarrow \infty$ so that the diffusion process is "stopped" across the region boundaries (i.e., edges) at locations of high gradients.

Two different diffusion functions $g(\cdot)$ have been proposed by Perona and Malik [12], which result in edge preserving filter defined as

$$g_1(\nabla I) = e^{(-(\|\nabla I\|/K)^2)}, \qquad (4)$$

$$g_2(\nabla I) = \frac{1}{1 + (\|\nabla I\|/K)^2}, \qquad (5)$$

where $K$ is a scale parameter (i.e., constant) to be tuned for a particular application. Perona and Malik [12] proposed that the value of $K$ can be fixed manually or using the "noise estimator" described by Canny [43]. In our algorithm fine details are separated using (4), which favors high contrast sharp transitions across multiexposure input series and the value of $K = 1/7$ was fixed manually based on experimentation.

The discrete formulation of Perona and Malik [12] anisotropic diffusion (i.e., base layer (B) in our case) is as given by

$$I_s^{t+1} = I_s^t \frac{\lambda}{|\eta_s|} \sum_{p \in \eta_s} g(\nabla I_{s,p}) \nabla I_{s,p}, \qquad (6)$$

where $I_s^t$ is a discrete version of input signal, $s$ determine the sample position in the discrete signal, and $t$ determines iterations. The constant $\lambda$ is a scalar that determines the rate of diffusion, $\eta_s$ represents the spatial neighborhoods of current sample position $s$, and $|\eta_s|$ is the number of neighbors.

To see the behavior of the Perona and Malik [12] filter at edges, we first analyze one-dimensional signal into base layer and detail layer. As can be seen in Figure 4, at base layer (i.e., the coarser level after diffusion), high-frequency textures disappear. The high texture details lost at the base layer are exactly reconstructed at the detail layer. However, detail layer is the difference between the input signal and the base layer, which is dominated by the large discontinuities characterized by the rapid oscillations (high-frequency variations) in the input signal. As a result, we are able to separate high texture details from edge transitions that are to be preserved during the fusion process. The continuous diffusive process for 1-D network structure (see Figure 3(a)) is as follows:

$$\frac{\partial I(x, t)}{\partial t} = \text{div}\left[g(\|\nabla I\|)\nabla I\right] \qquad (7)$$

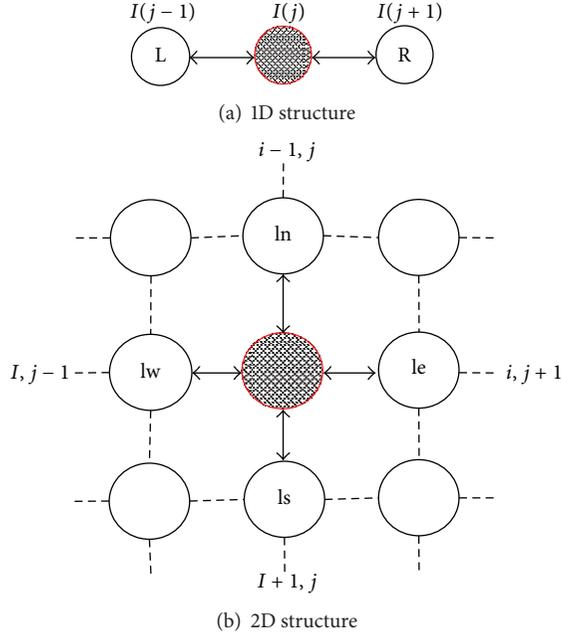

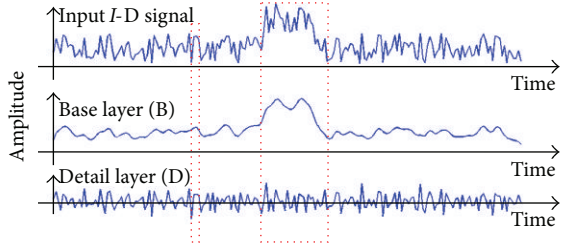

Figure 3: Gradient computation. (a) One-dimensional (1D) grid structure: the signal flow is calculated between two neighboring nodes (solid connection). (b) Two-dimensional (2D) grid structure: the signal flow is calculated between four neighboring nodes (solid connection).

Figure 4: Two layer decomposition of 1-dimensional signal based on Anisotropic Diffusion after 5 iterations with $k = 30$, $\lambda = 1/3$ and $|\eta_s| = 2$ (West & East). The 1-dimensional input signal ($I$) is decomposed into two main components: a low frequency base layer (B) and a high-frequency detail layer (D). Notice that the edges are preserved in the diffused image (i.e., base layer) and the detail layer yields fine details only.
and discrete formulation is written as

$$I_s^{t+1} = I_s^t + \frac{\lambda}{|\eta_s|}\left[g_L \cdot \nabla_L I + g_R \cdot \nabla_R I\right]_s^t, \quad (8)$$

where $I_s^t$ is a discrete version of input signal, $s$ determine the sample position in the discrete signal, and $t$ determines iteration. We found one iterations ($t = 1$) to be sufficient for the detail layer extraction across all of the input images we experimented at low computational time. The detailed analysis of effect of number of iteration on computational time and information present (i.e., entropy) in the fused image is given in Section 4. The constant $\lambda$ is a scalar that determines the rate of diffusion, $\eta_s$ represents the spatial neighborhood of current sample position $s$, the subscripts L and R depicting left and right, respectively, and $|\eta_s|$ is the number of neighbors (i.e., two in 1-D case), where $g_L$, and $g_R$ are the conduction coefficients across left and right spatial locations, respectively. The symbols $\nabla_L$ and $\nabla_R$ indicate the difference of left and right neighbor, respectively:

$$\nabla_L I_j \equiv I_{j-1} - I_j,$$
$$s = j \ (\text{sample location in 1-D grid}) \quad (9)$$

$$\nabla_R I_j \equiv I_{j+1} - I_j. \quad (10)$$

The anisotropic diffusion of two-dimensional grid shown in Figure 3(b) is given by the relation

$$\text{Base layer (B)} = I_s^{t+1}$$
$$= I_s^t + \frac{\lambda}{|\eta_s|}\left[g_N \cdot \nabla_N I + g_S \cdot \nabla_S I \right. \quad (11)$$
$$\left. + g_E \cdot \nabla_E I + g_W \cdot \nabla_W I\right]_s^t,$$

where $I_s^t$ is a discrete version of input image $I$, $s$ determine the pixel position in the discrete image, and $t$ determines iterations. The constant $\lambda$ is a scalar that determines the rate of diffusion, $\eta_s$ represents the spatial neighborhood of current pixel $s$ (North, South, East, and West), and $|\eta_s|$ is the number of neighbors (usually four), where $g_N$, $g_S$, $g_E$, and $g_W$ are the conduction coefficients across North, South, East, and West spatial locations. The symbols $\nabla_N$, $\nabla_S$, $\nabla_E$, and $\nabla_W$ indicate the difference of North, South, East, and West neighbor, respectively:

$$\nabla_N I_{i,j} \equiv I_{i-1,j} - I_{i,j},$$
$$\text{for } s = i, j \ (\text{pixel location in 2-D grid}), \quad (12)$$

$$\nabla_S I_{i,j} \equiv I_{i+1,j} - I_{i,j}, \quad (13)$$

$$\nabla_E I_{i,j} \equiv I_{i,j+1} - I_{i,j}, \quad (14)$$

$$\nabla_W I_{i,j} \equiv I_{i,j-1} - I_{i,j}, \quad (15)$$

$$\text{Detail layer (D)} \equiv I - B. \quad (16)$$

The base layer decomposition in (11) and detail layer decomposition in (16) of JUIT image are illustrated in Figure 5. From Figure 5, it can be visually seen that the base layer provides coarse details and the textures are almost eliminated. In Figure 6, we have illustrated the intensity profiles of base layers (blue color) and detail layers (red color) computed from multiexposure images. It is noticed that coarser and finer details are extracted across the visible details adaptively when the scene is captured with variable exposure times.

### 3.2. Weight Map Computation: Texture Filter Based on Local Range.
In the proposed algorithm, local range is used to generate weight map for nonuniform scaling to control contribution of pixels from base layers across all the multiple exposures. In Figure 7, we have illustrated that how local range is calculated in the range-filtered image from 3-by-3 neighborhood. This local range is likely to be very different from region to region in different images captured at variable exposure time. Well-exposed area will yield higher local range as compared to the overexposed and underexposed regions, which is illustrated in Figures 8 and 9. Local range is defined as follows:

$$R^*_{ij,k} = L_{max} - L_{min}, \qquad (17)$$

where $L_{max}$, $L_{min} \in$ local spatial window (i.e., 3-by-3) in the $k$th base layer

$$R_{ij,k} = \left[\sum_{k'}^{N} R^*_{ij,k'}\right]^{-1} R^*_{ij,k}, \qquad (18)$$

where $L_{max}$ and $L_{min}$ are the maximum and minimum values of the neighboring pixels within a 3-by-3 square window, respectively, and $R_{ij,k}$ (Normalized local range) is the weigh map at location $(i, j)$ in $k$th base image ($B_k$).

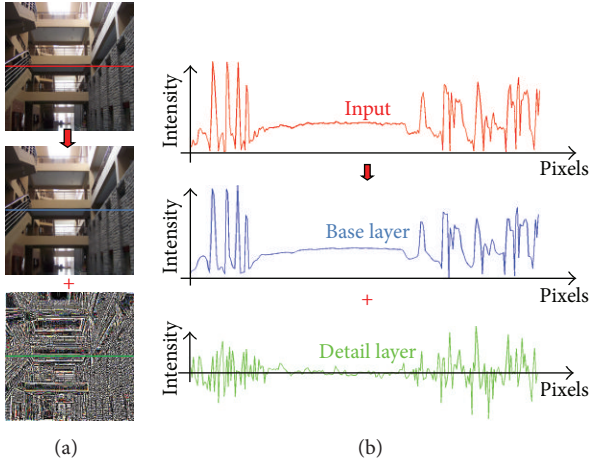

(a)  (b)

Figure 5: Two layer decomposition of two-dimensional signal (i.e., JUIT image) based on anisotropic diffusion (a) after five (5) iterations with $k = 30$, $\lambda = 1/7$, and $|\eta_s| = 4$ (region size of $3*3$ pixels). Intensity profiles (b) along a scan lines of two-dimensional input signal (red), base layer (blue), and detail layer (green). Notice that the strong edges are preserved in the diffused image (i.e., base layer) and the detail layer yields fine details only. Details compressed in the base layer are exactly reconstructed in detail layer.

It is commonly accepted that the higher the luminance variation region is the stronger the local range of that region to shield a pixel is. However, we find that the difference between the maximum and the minimum value of luminance also influences the probability of shielding the appropriate pixel. To compute such local range, our basic idea is illustrated in Figure 7. Then, the Gaussian pyramid of weight map is used to remove the influence of very high intensities and very low intensities present across the multiple exposures for producing the high-resolution image, which is described in Section 4. To illustrate the variation of local range in multiple exposures, we give four representative images as shown in Figures 9(a), 9(b), 9(c), and 9(d).

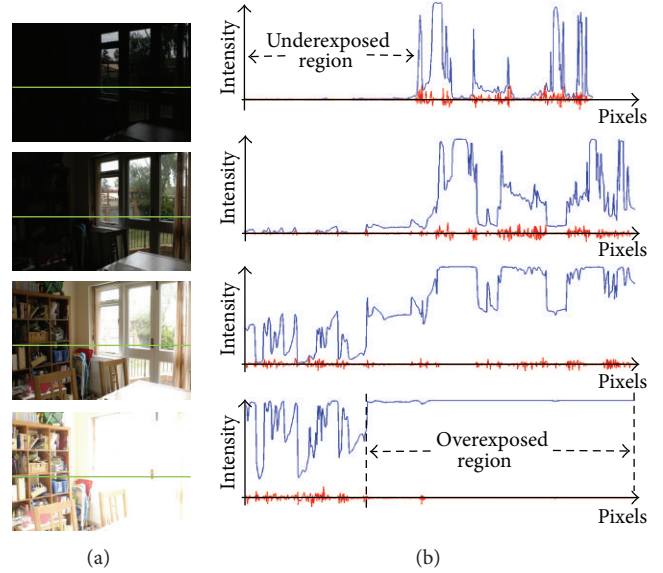

(a)  (b)

Figure 6: Intensity plots (b) along a scan lines of base layers (blue) and detail layers (red) obtained with the anisotropic diffusion after five iterations with $k = 30$, $\lambda = 1/7$, and $|\eta_s| = 4$ (region size of $3*3$ pixels) across all of the input exposures (a). Notice that coarser and finer details are extracted across the visible details adaptively when the scene is captured with variable exposure times.

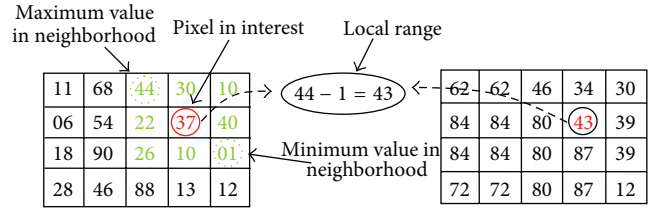

Figure 7: Illustration of local range calculation that is used as location adaptive weight map. The figure depicts how local range is calculated in the range-filtered image from 3-by-3 neighborhood. Green numbers are the neighbors considered to compute local range for the pixel of interest (i.e., displayed in red number).

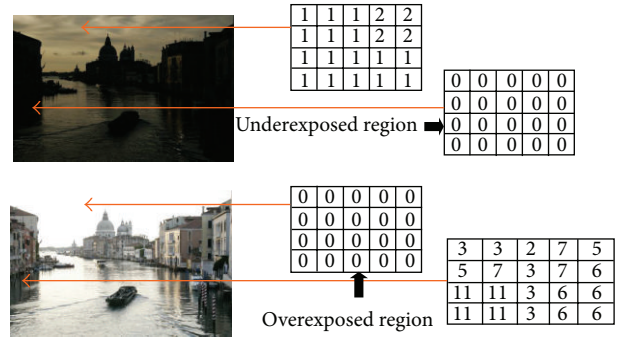

Figure 8: Effect of underexposed, normally exposed, and overexposed regions on local range. Note that local range will be different (i.e., zero for underexposed and overexposed regions) for the same region under different exposure values. The optimal window size for range calculation is 3-by-3 and the numerical values given in the box are calculated from the eight neighbors (see Figure 7).

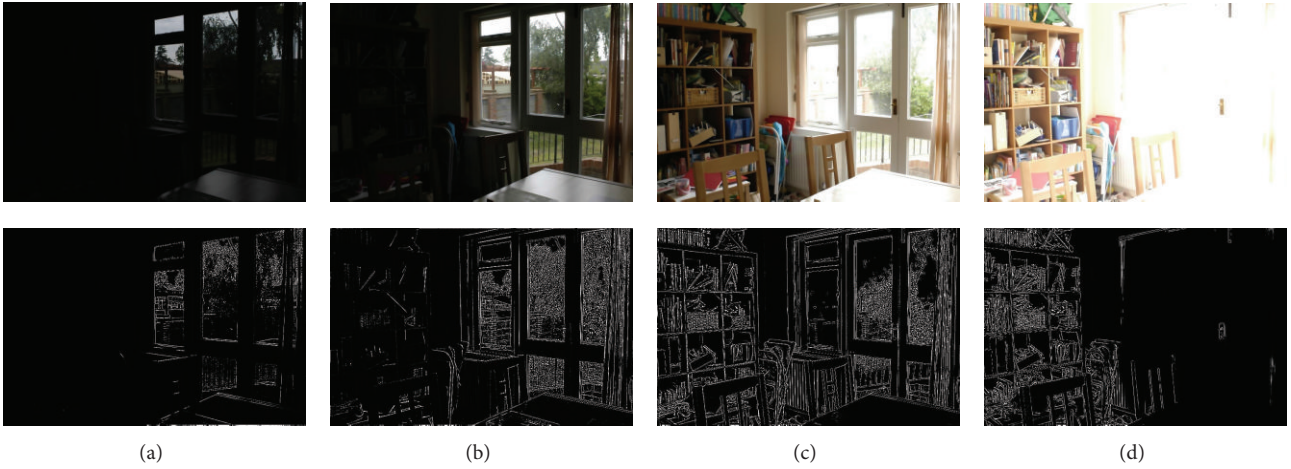

(a)      (b)      (c)      (d)

FIGURE 9: Illustration of local range analysis of base layers across the multiple exposures. The local ranges are varying with respect to the different exposure times. ((a), (b)) Base layers of underexposed images (top) and their corresponding texture features (bottom); (c) base layer of normally exposed image (top) and their corresponding texture features (bottom); (d) base layer of overexposed image (top) and their corresponding texture features (bottom). Note that well-exposed pixels have the brighter texture features (i.e., higher weights) across different exposure values. Input images: Jacques Joffre HDR Chief Photographer.

*3.3. Pyramid Generation and Construction of Fused Base Layers* ($B_f$) *across All Input Base Layers.* Researchers have attempted to synthesize and manipulate the features at several spatial resolutions [13, 44] that avoid the introduction of seam and artifacts such as contrast reversal or black halos. In the proposed algorithm, the bandpass components at different resolutions are manipulated based on texture features that determine the pixel value in the reconstructed fused base layer (B). We begin by constructing a Gaussian pyramid $GB_k^0, GB_k^1, \ldots, GB_k^d$ of $N$ input base layers across the input images, where $GB_k^0$ is the full resolution base layer and $GB_k^d$ is the coarsest level of the $k$th base layer in the pyramid. Lowpass filtering (convolving) a base layer $B_k$ with an equivalent weighting function and subsampled by removing every other pixel and every other row yields a Gaussian pyramid [11]:

$$GB_{ij,k}^l = \sum_{m=-2}^{2} \sum_{n=-2}^{2} w(m,n) GB_k^{l-1}(2i+m, 2j+n). \quad (19)$$

Here $l$ ($0 < l < d$) refers to the number of levels in the pyramid and $k$ ($1 < k < N$) refers to the number of input base layers and $w(m, n)$ is an equivalent weighting function. In our case, the Gaussian pyramid is generated with $a = 0.4$ [11], which yields more Gaussian-like equivalent weighting functions. A Laplacian pyramid of input base layers $LB_k^0, LB_k^1, \ldots, LB_k^d$ is created containing band-pass images of decreasing size and spatial frequency:

$$LB_k^l = GB_k^l - \text{EXPAND}\left(GB_k^{l+1}\right), \quad k = 1, \ldots, N, \ l = 0, \ldots, d, \quad (20)$$

$$LB_k^d = \text{EXPAND}\left(GB_k^d\right), \quad (21)$$

where the expanded image $GB_k^{l+1}$ is the same size as the $GB_k^{l-1}$, and $LB_k^l$ is the level of Laplacian pyramid of $k$th base image. Each Laplacian level contains local spatial information at increasing fine details.

The patches extracted from the input base layers are used for texture analysis (i.e., local range). We calculate a weight around every pixel within a 3-by-3 window. The value of the weighting function for each pixel depends on the maximum and minimum intensity value of the neighbors within the window. Next, the local range calculated from base layer (i.e., diffused image) in (11) is computed in top-down fashion, similar to that described in [11]:

$$GR_{ij,k}^l = \sum_{m=-2}^{2} \sum_{n=-2}^{2} w(m,n) GR_k^{l-1}(2i+m, 2j+n), \quad (22)$$

where $l$ and $k$ denote the level of Gaussian pyramid of local range $GR_k^0, GR_k^1, \ldots, GR_k^d$ of $k$th input range filtered image with $GR_k^0$ as the full resolution image and $GR_k^d$ is the coarsest level in the pyramid.

Gaussian pyramid of texture feature (i.e., local range) acts as weight map that determines the contribution of pixels from the base layers across all of the multiple exposures. The Laplacian pyramid of base layer $LB_k^l$ multiplied with the corresponding Gaussian pyramid of texture feature $GR_k^l$ and summing over $k$ yields modified Laplacian pyramid $L^l$:

$$L_{ij}^l = \sum_{k=1}^{N} LB_{ij,k}^l GR_{ij,k}^l. \quad (23)$$

In the case of the image averaging, the output pixels are an average of input pixel's luminance values, which reduce noise in the final image, but the contrast of details is compromised and the images can look washed out. Note, however, we have found that Pyramid fusion [23] performs very well on base layer fusion when modified with weight maps giving more pleasing results with optimal contrast enhancement.

The fused base layer that contains well-exposed pixels is reconstructed from $L^l$ by expanding each level and summing

$$B_f = L^0 + L^1 + L^2 + \cdots + L^l. \tag{24}$$

We found that the modification of Laplacian pyramid in top-down fashion eliminates underexposed and overexposed regions in the fused base layer that leads to well-exposed image without the introduction artifacts. See Figure 10 for an illustration of the proposed idea.

### 3.4. Construction of Fused Detail Layer ($D_f$) and Detail Layer Enhancement.
The detail layers $D_1, D_2, \ldots, D_N$ computed from (1) contain the smaller changes in intensity. There are mainly three parameters that control the behavior of base layer and detail layer computation in our exposure fusion approach. Referring to (11), $t$ and constant $\lambda$ determine the iterations and the rate of diffusion, respectively. The constant value $K$ can be chosen manually or by using the "noise estimator" proposed by Perona and Malik [12]. As a consequence, we can vary these three parameters to moderate texture details in the fused image. When $t$ increases, adjacent pixels with large intensity differences are ignored (i.e., more smoothing at edges), which leads to larger details in the residual layers across different exposures. However, if $t$ becomes too small, fewer details are preserved in the residual layers across all of the input exposures with smaller computational time. In order to balance the computational time and detail in the fused image, we have fixed and suggested $t = 1$, $\lambda = 1/7$, and $K = 30$ in all experiments, which reveals reasonably good results. More detailed analysis of effects of these free parameters is given in Section 4. We have presented two alternative options for constructing the residual image (i.e., detail layer $D_f$) and manipulating the details in the fused image. We believe that both options can be utilized, depending on the application.

#### 3.4.1. User Driven.
In order to compute residual layer having rich texture detail, we use a weighting factor $\alpha_1$ determined by the user (typically 1.2 in our approach, see Figure 1(b)) and the residual layer is obtained as a linear combination of the detail layers across the input multiexposure images:

$$D_f = \frac{\sum_{k=1}^{N} \alpha_1 D_k}{N}. \tag{25}$$

This straightforward option allows the user to control the contribution of texture details directly from the input detail layers across all of the input images. We found that this simple technique is effective to boost weak details in the fused image but yield overenhancement at the strong edges. Furthermore, to manipulate detail layers across all of the input images precisely, we present a second technique that enhances weak details, while avoids artifacts near the edges.

#### 3.4.2. Sigmoid Function Based Detail Layer Manipulation and Fusion.
The second alternative option to enhance fine details in the fused image is based on monotonic nonlinear activation function, where the resultant residual layer is computed as follows:

$$D_f = \sum_{i}^{n'} \sum_{j}^{m'} \frac{\alpha_2 S\left(D_k^{(i,j)}\right)}{N}, \quad k = 1, \ldots, N, \tag{26}$$

where $\alpha_2$ is a fixed weight ($\alpha_2 = 2$ is found to be suitable in our approach in most cases) and $S(\cdot)$ is the 1-dimensional sigmoid function

$$S(t) = \frac{1}{1 + e^{-at}}, \tag{27}$$

where $t \in \mathcal{R}$ is the independent variable and $a \in \mathcal{R}$ is a weight parameter of the sigmoid function. Figure 11 shows a 1-dimensional sigmoid with different weight values. The weight parameter used in our approach was set to 27.

Let $\theta$ be a fixed threshold to further control the sharpness of sigmoid function, which is manually chosen by the operator. The 1-dimensional sigmoid function with threshold $\theta$ is given by

$$S(t) = \frac{1}{1 + e^{(-at+\theta)}}. \tag{28}$$

In our approach $\theta$ is responsible for global contrast management. The detailed analysis of selection of these parameters is given in Section 4. Minai and Williams [45] have presented the sigmoid with threshold as a neuron activation function in artificial neural networks and recurrence relations for calculating derivatives of any order. The first derivative of sigmoid function $S'(t)$ is computed as

$$S'(t) = aS(t)(1 - S(t)). \tag{29}$$

## 4. Experimental Results and Analysis

### 4.1. Comparison with Other Exposure Fusion, Multifocus Fusion, and Tone Mapping Methods.
In this paper, we have implemented our algorithm in MATLAB-7.5.0 and run on a PC with 2.2 GHz i5 processor and 2 GB of RAM. As shown in Figures 12(b) and 13(b), note that the fused image provides natural contrast and has no noticeable artifacts. We tested our proposed algorithm on a variety of bracketed sequences. The proposed approach is computationally simple and results are comparable to several tone mapping algorithms. Figure 2 shows the block diagram of the proposed texture feature based detail enhancing exposure fusion technique.

Figures 12, 13, 14, and 15 show the comparison of the proposed experimental results. In these experiments, optimal block size for weight map calculation was 3-by-3. Figures 12(a) and 13(a) show image pairs of the "igloo" and "door" image sequence (size of $221 \times 336 \times 6$ and $223 \times 332 \times 6$, resp.). We can see from Figure 12(b) that all the light in the scene that appears to come from natural light source is optimally reproduced with crisp shadows. In Figure 12, one auto-exposure image captured with the digital camera and two

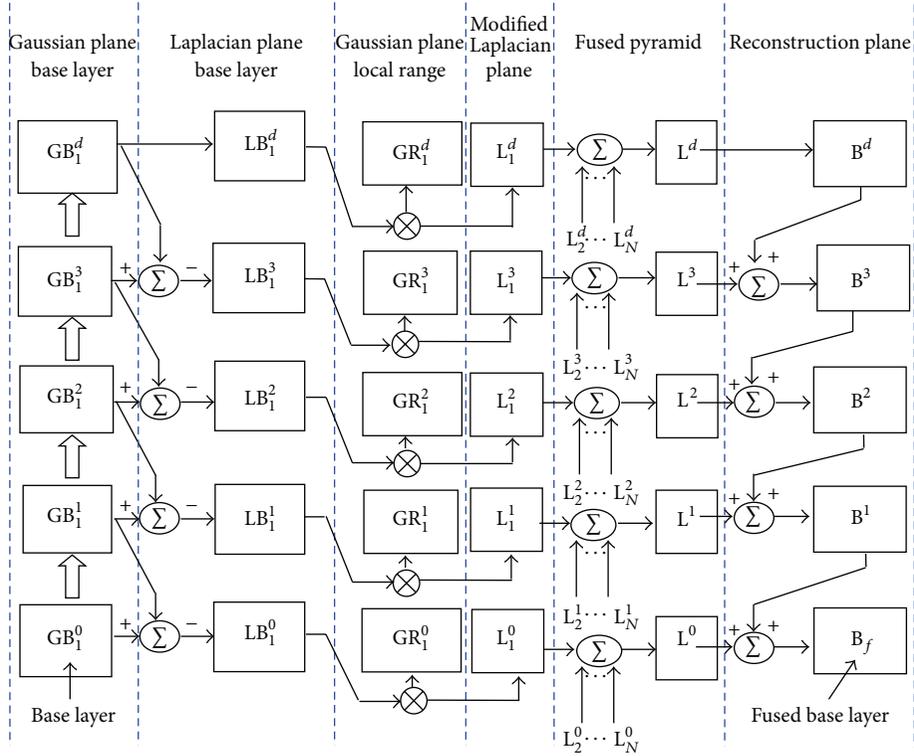

Figure 10: Base layer manipulation and fusion: illustrating the conceptual framework of texture feature (i.e., decision map) based pyramid fusion approach of coarser detail across input images. Note that for the concept simplicity, here we have generated the Laplacian pyramid of single base layer and the Gaussian pyramid of the corresponding texture features, where $\tilde{L}_2^0, \tilde{L}_3^0, \ldots, \tilde{L}_N^0$ are the modified Laplacian pyramid of base layers across all of the multiexposures.

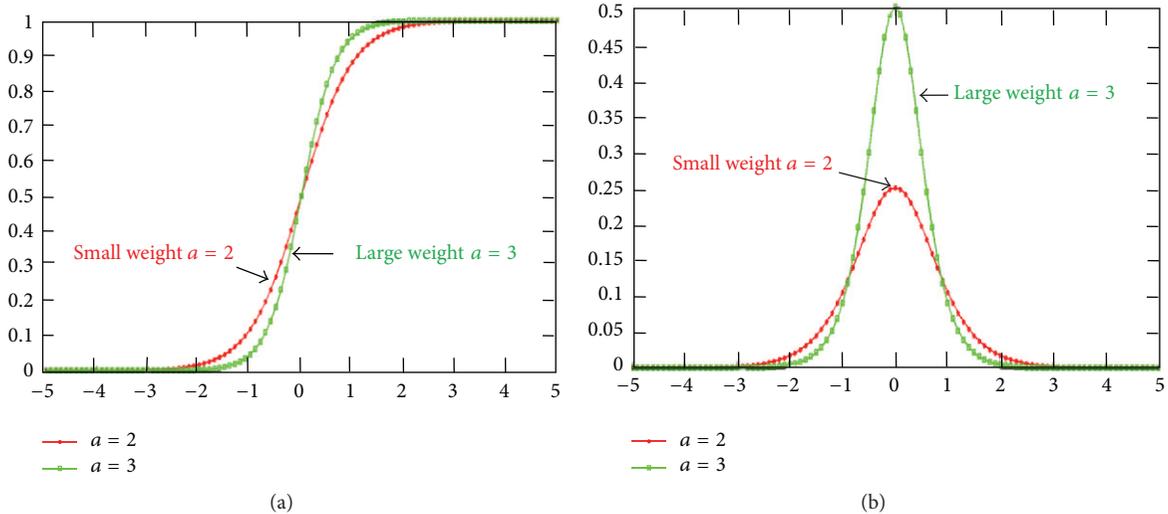

Figure 11: The effect of weight (i.e., $a$) on sigmoid function and derivative of sigmoid function. (a) The sharpness of the sigmoid in (28) varies according to the value of weight. With larger value of $a$, the sigmoid becomes a threshold function. (b) The first derivative of sigmoid function in (29) for $a = 2$ and $a = 3$.

recently proposed fusion results of "igloo" are demonstrated. It can be noticed that the proposed technique provides better texture details in highlights and shadows as compare to the results of autoexposure (Figure 12(c)) and Mertens et al. [13] (Figure 12(d)). It may also be observed that the brightly illuminated region (i.e., sky area) is overexposed in the result proposed by Shen et al. [46] (see Figure 12(e)). Figure 13(b) shows more comparison example of our result for scene depicting outdoor and indoor details. The proposed techniques is visually compared with the results of autoexposure (Figure 13(c)), and recently proposed Mertens et al. [13] (Figure 13(d)) and Zhang and Cham [47] (Figure 13(e)). By contrast, it is seen that our method combines the best of multiple exposures into one realistic-looking image that is much closer to what our eyes originally saw. However, both indoor and outdoor details of input LDR images (Figure 13(a)) are simultaneously produced in the fused image with optimal contrast and without the introduction of artifacts. Although Mertens et al. [13] have produced comparable results, it does not preserve all details from input LDR images. As shown in Figure 13(e), the results produced by Zhang and Cham [47] depict washed out details in underexposed regions which are not able to preserve texture details from input LDR shots.

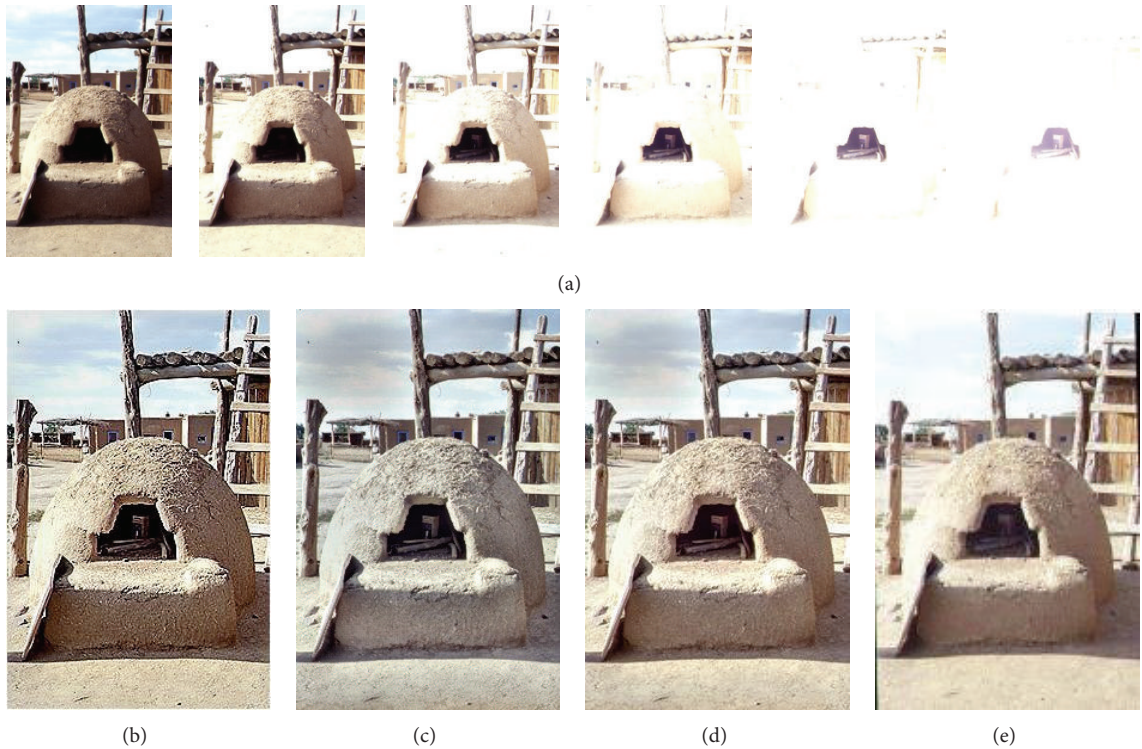

Figure 12: (a) Images representing multiple exposures of an outdoor scene depicting highlights from natural sun light and shadows. (b) Our technique fuses the multiple exposures to obtain high quality image. Note that the fused image yields more texture details and natural contrast without the introduction of artifacts. (c) Autoexposure, (d) Mertens et al. [13], and (e) Shen et al. [46]. Input images are courtesy of Shree Nayar.

To further compare our results visually with Mertens et al. [13] and Shen et al. [46], respectively, Figures 14(a), 14(b), and 14(c) depict a close-up view. The first row of Figure 1 depicts the "house" LDR image sequence of size $752 \times 500 \times 4$ which is provided by Mertens et al. [13]. It can be observed that the texture details (see the fine textures on the chair and books behind the chair) are accurately preserved in the proposed fused image (see Figure 14(a)).

In this section, we compare our results for "Belgium house" image sequence of size $1025 \times 769 \times 9$ (see Figure 15(a)) with the popular exposure fusion and tone-mapped HDR images, which are depicted in Figures 15(b), 15(c), 15(d), 15(e), and 15(f). In particular, we do compare our results with the perceptually driven works [16] and low curvature image simplifier (LCIS) hierarchical decomposition [48]. As shown in Figure 15(b), our technique yields fine texture details in the fused image with natural contrast that is entirely free of halo artifacts. To illustrate the effectiveness of proposed approach, we illustrate close-up comparison in Figures 15(b)–15(f). Larson et al. [16] presented a dynamic range compression method based on a human visual system adaptation, and it was also found to suffer from halo artifacts and does not offer good color information (see Figure 15(e)). Tumblin and Turk [48] preserve fine details in the image, while weak halo artifacts are present around certain edges in strongly compressed areas (see Figure 15(f)). Experimental results have demonstrated that proposed method worked very well on a variety of multiple exposures and preserved the original scene's relative visual contrast impression.

Furthermore, to check the effectiveness of the proposed algorithm for other applications, we have employed the same technique for the fusion of multifocus image series (Figures 16, 17, and 18) and images captured with flash and no-flash (Figure 19). Figure 16(a) illustrates two partially focused RGB images (focused on two different targets). It is illustrated in Figure 16(b) that the color information is preserved in the fused image with better visualization of texture details. On the other hand we have tested and compared our approach for two sets of multifocused gray scale images of "table" and "clock", which are illustrated in Figures 17(a)–17(d) and Figures 18(a)–18(d), respectively. As demonstrated in

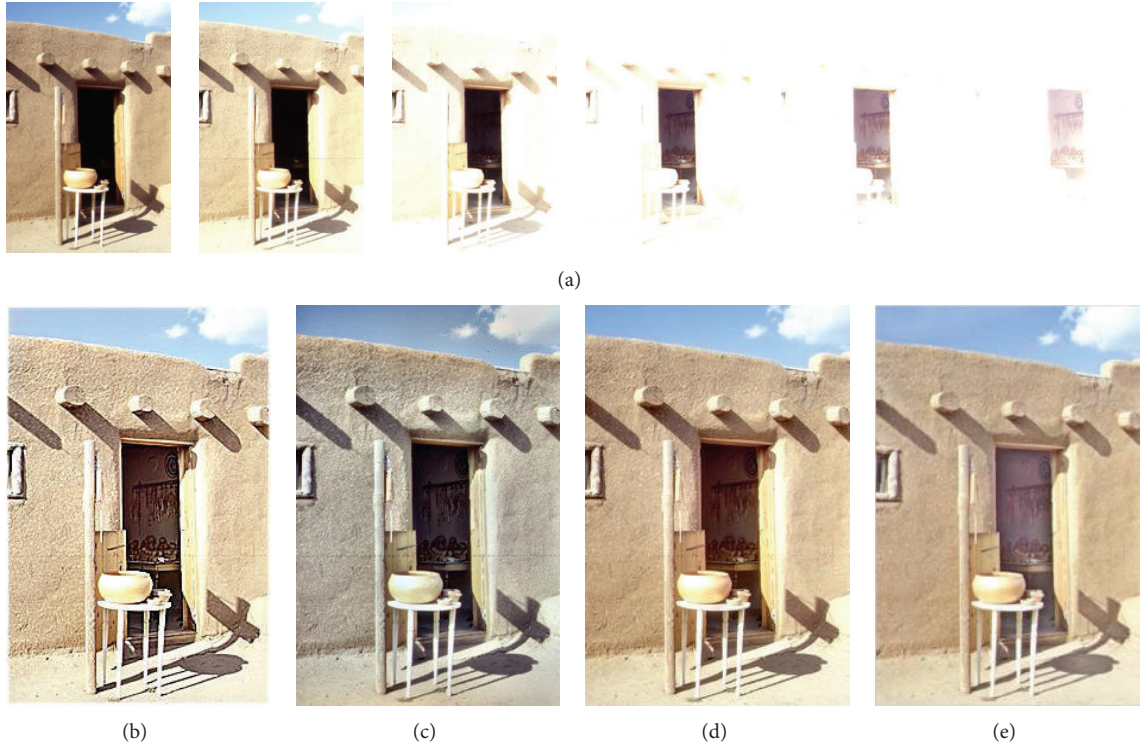

Figure 13: (a) Images representing multiple exposures of an indoor and outdoor scene depicting sunlit details and shadows. (b) Our technique fuses the multiple exposures to obtain high quality image. Note that the fused image yields more texture details and natural contrast without the introduction of artifacts. (c) Autoexposure, (d) Mertens et al. [13], and (e) Zhang and Cham [47]. Input images are courtesy of Shree Nayar.

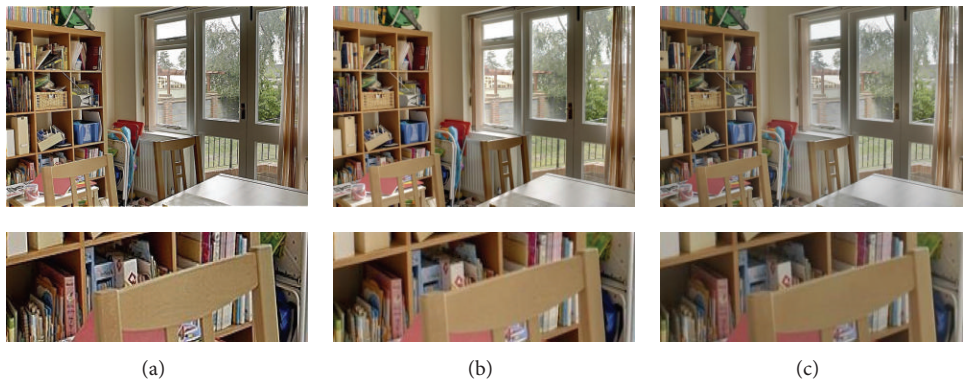

Figure 14: House: comparison results to other recent exposure fusion techniques. (a) Results of our new exposure fusion method, (b) Mertens et al. [13], and (c) Shen et al. [46]. Note that our method yields enhanced texture and edge features. Input image sequence is courtesy of Tom Mertens.

Figure 17(c) that our results produce pleasing image with rich texture details, the results produced by P. Hodáková et al. [49] in Figure 17(d) do not reveal fine details present across all input images. It can easily be noticed that our fused image in Figure 18(c) extracts more information from the original images. Moreover, Adu and Wang's technique [27] in Figure 18(d) appears washed out, which is responsible for losing perception of fine texture details.

Finally, we have also tested our technique on two sets of images captured with flash and no-flash images (see Figures 19(a) and 19(b)). Our approach provides interesting solution for fusing the flash/no-flash image pair. Figure 19(c) illustrates our results, which combine details from the flash/no-flash image pair. As shown in Figure 19(c), the proposed approach allows removal of highlights from flash images and yields high quality flash image with optimal contrast and detail enhancement. The experimental results in Figure 19(c) depict largest amount of information and has relatively better contrast than that of results of Mertens et al. [13] in Figure 19(d).

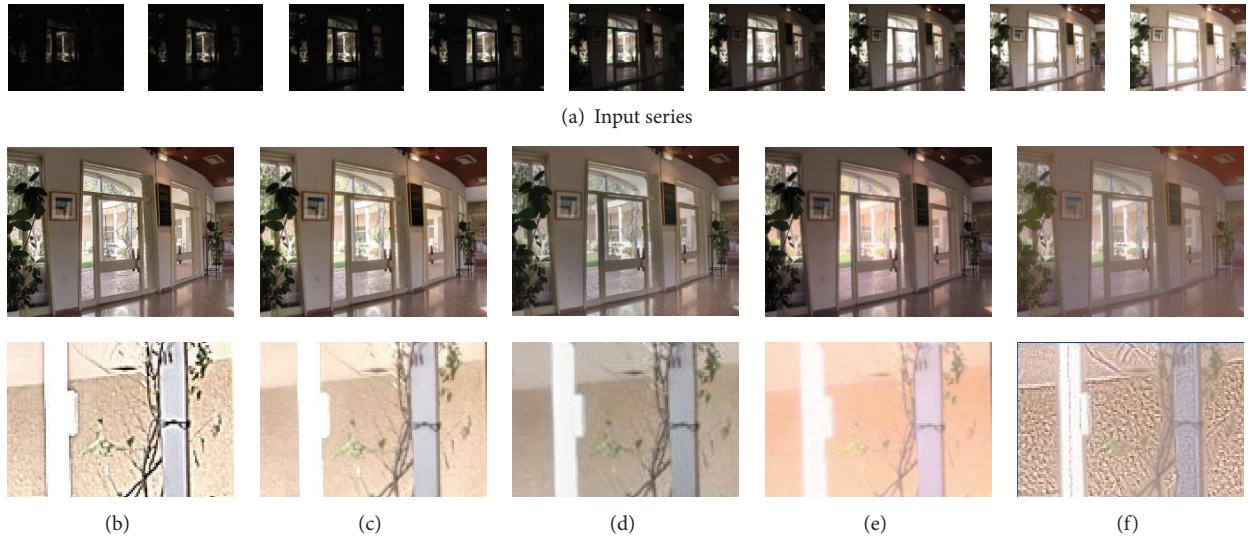

Figure 15: Belgium house: (a) series of multiple exposures depicting both indoor and outdoor areas. The exposure value is varying from (1/1000 of a second) to (1/4 of a second). Comparison results to other popular tone mapping techniques. (b) Results of our new exposure fusion method, window size = 3-by-3, (c) Mertens et al. [13], (d) Shen et al. [46], (e) results of Larson et al. [16], (f) results of LCIS method [48]. Note that our method yields combined features that can only be recorded using different exposures. Input images are courtesy of Dani Lischinski.

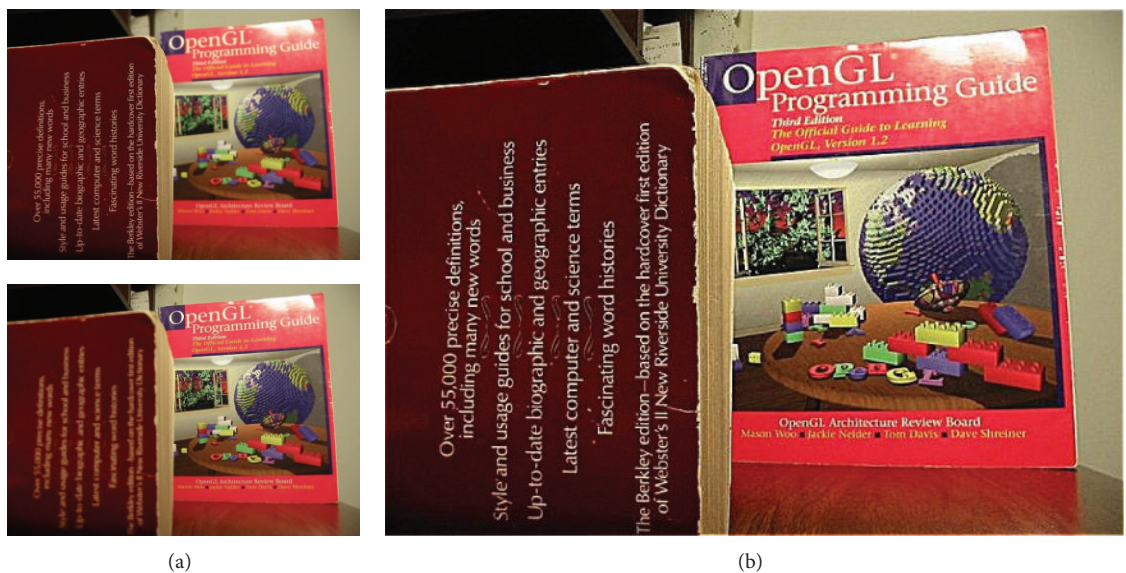

Figure 16: Book image. (a) Two partially focused images (focused on different targets) and (b) image generated by the proposed approach, which illustrate that the fused image extracts more color and texture details from the original input images (input sequence is courtesy of Adu and Wang [27]).

To perform visual inspection of exposure fusion results of Mertens et al. [13] shown in Figures 12(d), 13(d), 14(b), 15(c), and 19(d) are produced with the help of Matlab code provided by the authors. The original results of generalized random walks based fusion [46] in Figures 12(e), 14(c), and 15(d) are provided by the authors on request. All the experimental results of Zhang and Cham [47] in Figure 13(e), tone mapped HDR [16, 48] in Figures 15(e) and 15(f), and multifocused fusion [27, 49] in Figures 17(d) and 18(d) are taken from its papers. It is noticed that unlike the previous work such as [46], our approach preserves more details with higher contrast and does not require further postprocessing. Thus, this approach can be utilized in computer graphics applications.

4.2. Analysis of Free Parameters. To analyze the effect of iteration on quality score [50], entropy, computational time and mean square error (MSE), we have illustrated four plots (see Figures 20(a), 20(b), 20(c), and 20(d), resp.) at a different

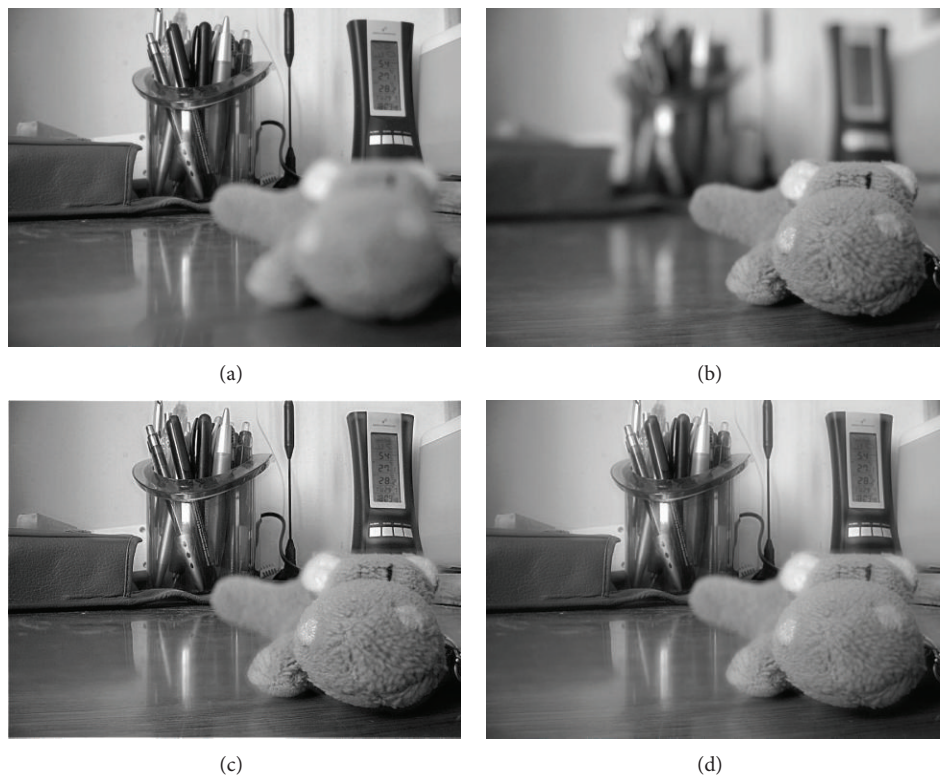

Figure 17: Table image. ((a), (b)) Two partially focused images (focused on different targets), (c) image generated by the proposed approach, which illustrates that the fused image extracts more information from the original images, and (d) Hodáková et al. [49].

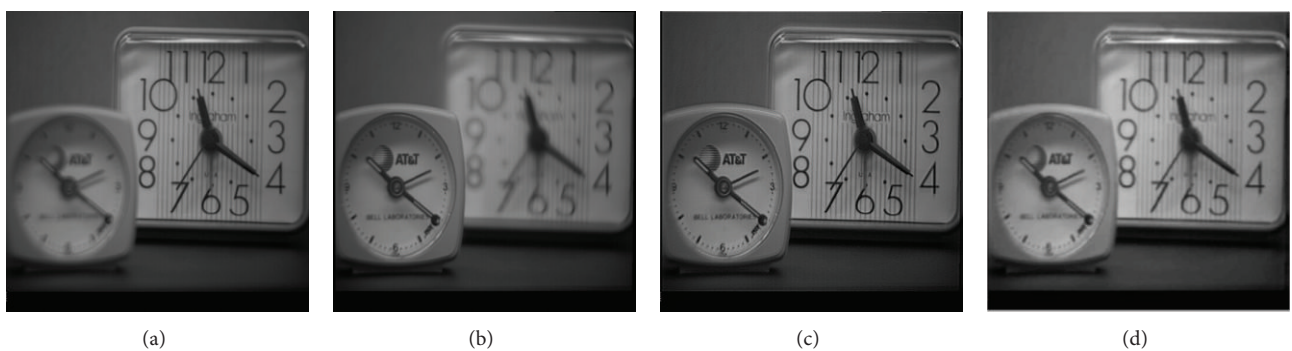

Figure 18: Clock Image. ((a), (b)) Two partially focused images (focused on different targets), (c) image generated by the proposed approach, which illustrates that the fused image extracts more information from the original images, and (d) Adu and Wang [27] (input sequence is courtesy of Adu and Wang [27]).

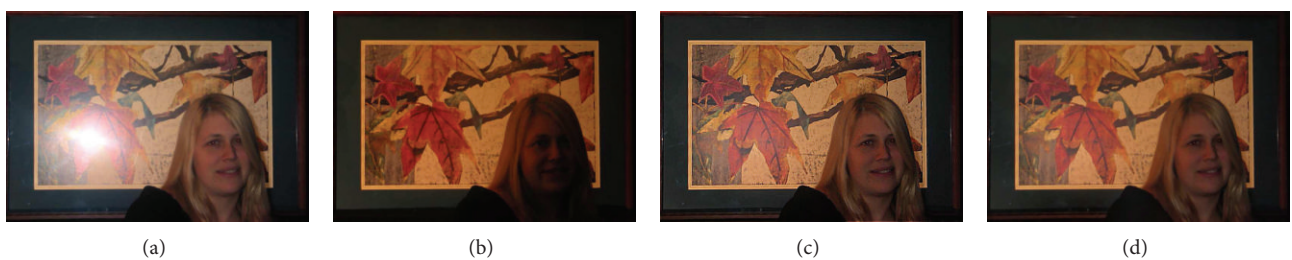

Figure 19: ((a), (b)) Input images photographed with and without flash; (c) enhanced fused image by proposed algorithm which maintains the warm appearance and the sharp details after removing strong highlight, and (d) Mertens et al. [13]. Images taken from Agrawal et al. [24].

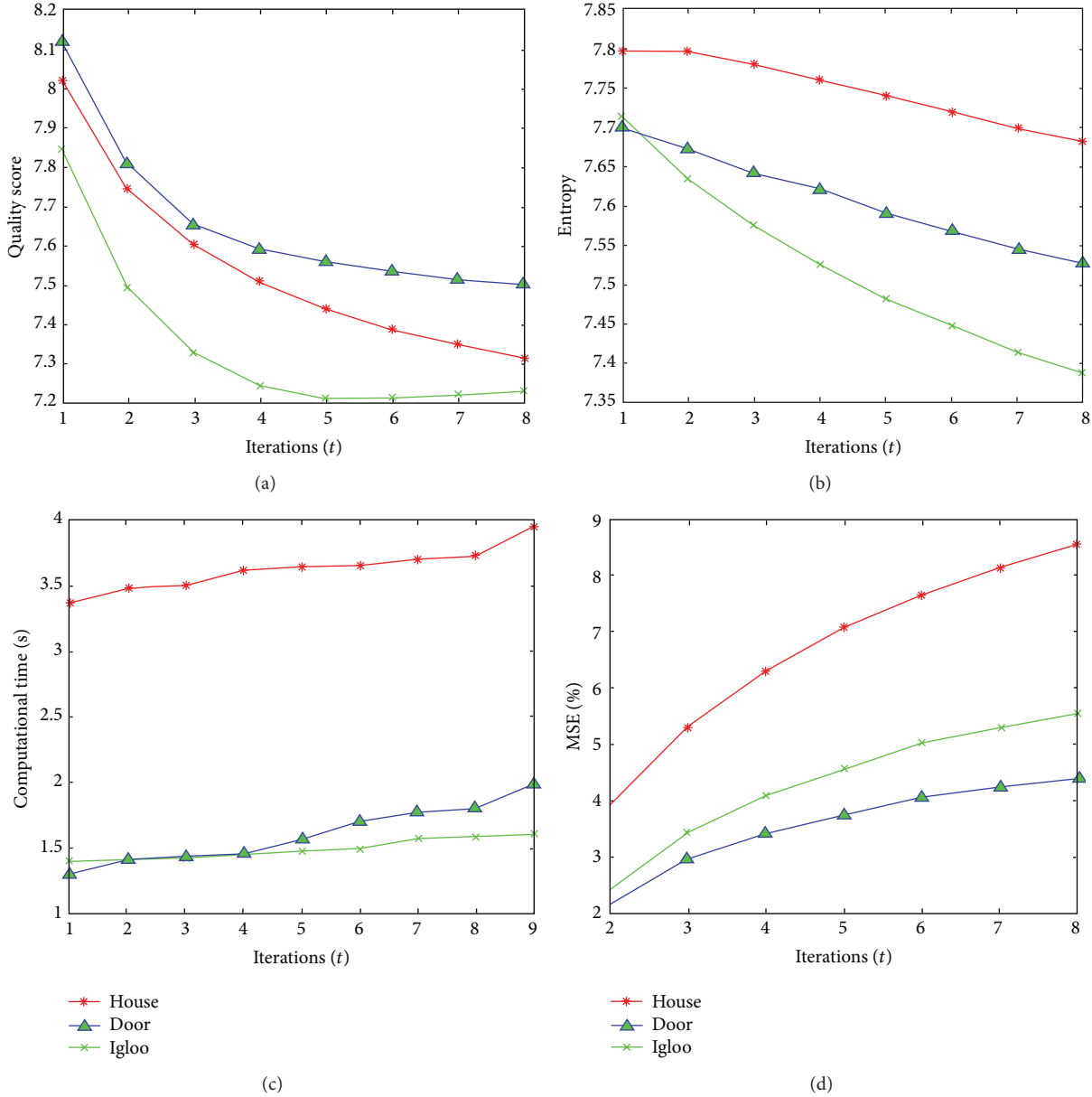

FIGURE 20: Analysis of number of iterations used for base layer computation. Mean square error is defined as the relative difference from the results generated with $t = 1$. Maximum quality score and entropy are only observed when $t = 1$. It is observed that MSE and computational time increase as $t$ increases. (a) Effectiveness of $t$ on quality score, (b) entropy, (c) computational time, and (d) error introduced.

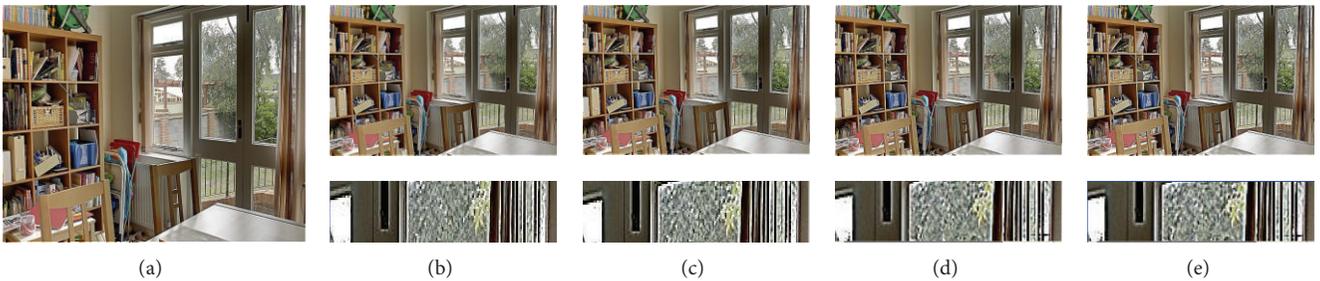

FIGURE 21: House image. The free parameter $t$ in (11) is used to control detail enhancement. We have found that $t = 1$ is sufficient for fine details extraction and gives better results for most cases. Higher value of $t$ brings in artifacts near strong edges. (a) $t = 1$, (b) $t = 2$, (c) $t = 3$, (d) $t = 4$, and (e) $t = 5$.

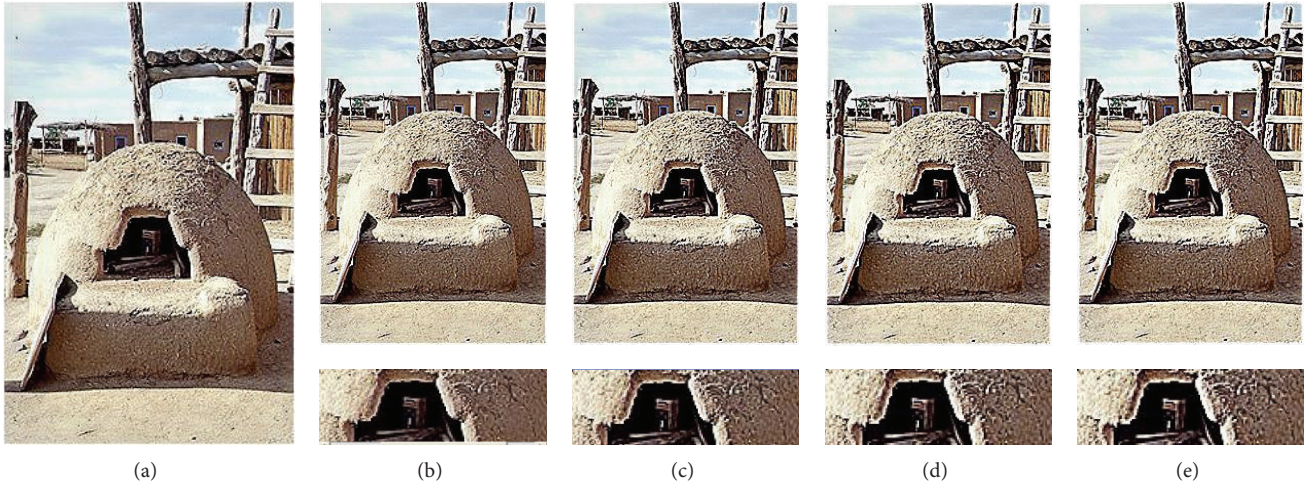

Figure 22: Igloo image. The free parameter $t$ in (11) is used to control detail enhancement. We have found that $t = 1$ is sufficient for fine details extraction and gives better results for most cases. Higher value of $t$ brings in more details while introducing artifacts near strong edges. (a) $t = 1$, (b) $t = 2$, (c) $t = 3$, (d) $t = 4$, and (e) $t = 5$.

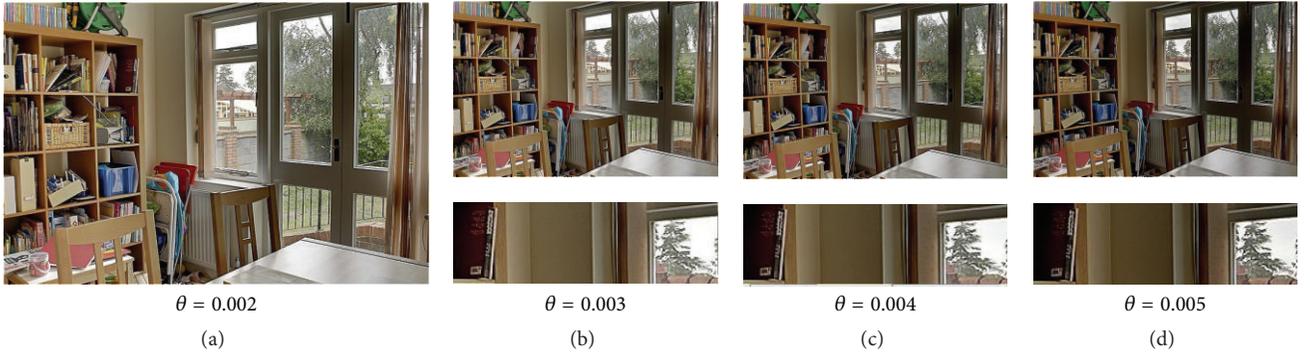

Figure 23: House image. The free parameter $\theta$ in equation is used to control sharpening. We have found that $\theta = .002$ gives better results for most cases. Higher value of $\theta$ brings in more details in highly illuminated areas. (a) $\theta = .002$, (b) $\theta = .003$, (c) $\theta = .004$, and (d) $\theta = .005$.

value of iteration ($t$) for input image sequences of "house," "igloo," and "door." To assess the effect of iteration on fusion performance, the quality score [50] and entropy were adopted in all experiments. To measure computational time, all the experiments were executed on a PC with 2.2 GHz i5 processor and 2 GB of RAM. The MSE is estimated as the difference between pixel values implied by different iterations (i.e., $t = 2, 3, 4, 5, 6, 7, 8$) and the reference image obtained with low iteration value (i.e., $t = 1$). We fix $\lambda = 1/7$, $K = 30$, $\alpha = 2$, and $a = 27$ in all experiments and they are set as default parameters.

First, to analyze the effect of iteration on quality score, entropy, and computational time the threshold ($\theta$) used in (29) for scale selection was set to 0.002. As shown in Figures 20(a) and 20(b), the best fusion performance is given at $t = 1$. The quality score and entropy decrease as $t$ increases. As shown in Figure 20(c), the computational time increases as $t$ increases. The visual inspection of effect of $t$ on image sequences (i.e., "house" and "igloo") is depicted in Figures 21 and 22, respectively. It can easily be noticed from the close up view (see Figures 21(b), 21(c), 21(d), 22(b), 22(c), and 22(d)) that as $t$ increases, the sharp edges get brighter and therefore lead to artifacts at sharp edges. To analyze the error (i.e., MSE) introduction against $t$ the one of the image produced with $t = 1$, $\lambda = 1/7$, $K = 30$, $\alpha = 2$, $\theta = .002$, and $a = 27$ is considered as reference image. The error increases as the number of iterations ($t$) increases. From Figure 20(d), it can also be noticed that when $t = 8$, the total error introduced is still less than 9%.

In the analysis of threshold ($\theta$), we fix $t = 1$, $\lambda = 1/7$, $K = 30$, $\alpha = 2$, and $a = 27$. Four results obtained by different $\theta$'s are shown in Figures 23(a), 23(b), 23(c), and 23(d). For the result in Figure 23(a), the value of $\theta$ is .002, and in Figures 9(b)–9(d) the values of $\theta$ are .003, .004, and .005. Increasing the value of $\theta$ for controlling the sharpness of sigmoid function reveals more details in strongly illuminated areas (i.e., overexposed regions) and the image gets darker. In order to balance the details and contrast, we have found that $\theta = .002$ generates reasonably good results for all cases. Finally, from these experiments, we have concluded that the best results were obtained with $t = 1$, $\lambda = 1/7$, $K = 30$, $\alpha = 2$, $\theta = .002$, and $a = 27$, which yield more details and good contrast.

## 5. Conclusions

In this paper, we have proposed texture features based exposure fusion, which has applicability to preserve the details in poorly and brightly illuminated regions. Our method uses texture features to modify Laplacian pyramid of the base layer across multiple exposures at different spatial scales and then constructs a well-exposed low dynamic range image by expanding, then summing all the levels of the fused Laplacian pyramid for the different base layers. Nonlinear diffusion filters based on partial differential equations (PDE) were proposed to preserve fine details. Experimental results demonstrated that our approach has applicability for other applications, including multifocus image fusion and fusion of flash/no-flash image pairs, in which the fine details are preserved accurately. In particular, the main contribution of our work is proposal of a novel technique that fuses details in edge preserving manner from images captured at variable exposure settings without the introduction of artifacts. In future, we will explore the applicability of single resolution techniques to reduce the computational cost of the proposed exposure fusion algorithm.

## Acknowledgments

The authors would like to thank Jacques Joffre, Dani Lischinski, Shree Nayar, Jack Tumblin, and Greg Ward for the permission to use their images. They would like to thank Rui Shen for providing images for analysis purpose. They are also thankful to the reviewers for their valuable suggestions and proposed corrections to improve the quality of the paper.

## References

[1] E. Reinhard, G. Ward, S. Pattanaik, and P. Debvec, *High Dynamic Range Imaging Acquisition, Manipulation, and Display*, Morgan Kaufmann, 2005.

[2] J. A. Ferwerda, S. N. Pattanaik, P. Shirley, and D. P. Greenberg, "Model of visual adaptation for realistic image synthesis," in *Proceedings of the Computer Graphics Conference (SIGGRAPH '96)*, pp. 249–258, August 1996.

[3] P. E. Debevec and J. Malik, "Recovering high dynamic range radiance maps from photographs," in *Proceedings of the 24th ACM Annual Conference on Computer Graphics and Interactive techniques (SIGGRAPH '97)*, pp. 369–378, Los Angeles, Calif, USA, August 1997.

[4] S. Mann and R. W. Picard, "Being "undigital" with digital cameras: extending dynamic range by combining differently exposed pictures," in *Proceedings of the IS&T's 48th Annual Conference*, pp. 442–448, May 1995.

[5] K. Jacobs, C. Loscos, and G. Ward, "Automatic high-dynamic range image generation for dynamic scenes," *IEEE Computer Graphics and Applications*, vol. 28, no. 2, pp. 84–93, 2008.

[6] G. Ward, "Fast, robust image registration for compositing high dynamic range photographs from hand-held exposures," *Journal of Graphics Tools*, vol. 8, no. 2, pp. 17–30, 2003.

[7] A. Tomaszewska and R. Mantiuk, "Image registration for multi-exposure high dynamic range image acquisition," in *Proceedings of the International Conference on Computer Graphics, Visualization and Computer Vision*, Plzen, Czech Republic, 2007.

[8] E. Reinhard, M. Stark, P. Shirley, and J. Ferwerda, "Photographic tone reproduction for digital images," *ACM Transactions on Graphics*, vol. 21, no. 3, pp. 267–276, 2002.

[9] H. Seetzen, W. Heidrich, W. Stuerzlinger et al., "High dynamic range display system," *ACM Transaction on Graphics*, vol. 23, no. 3, pp. 760–768, 2004.

[10] H. Seetzen, L. A. Whitehead, and G. Ward, "A high dynamic range display using low and high resolution modulator," in *Proceedings of the Society for Information Display International Symposium*, vol. 34, pp. 1450–1453, 2003.

[11] P. J. Burt and E. H. Adelson, "The Laplacian pyramid as a compact image code," *IEEE Transactions on Communications*, vol. 31, no. 4, pp. 532–540, 1983.

[12] P. Perona and J. Malik, "Scale-space and edge detection using anisotropic diffusion," *IEEE Transactions on Pattern Analysis and Machine Intelligence*, vol. 12, no. 7, pp. 629–639, 1990.

[13] T. Mertens, J. Kautz, and F. Van Reeth, "Exposure fusion: a simple and practical alternative to high dynamic range photography," *Computer Graphics Forum*, vol. 28, no. 1, pp. 161–171, 2009.

[14] R. Fattal, M. Agrawala, and S. Rusinkiewicz, "Multiscale shape and detail enhancement from multi-light image collections," in *Proceedings of the International Conference on Computer Graphics and Interactive Techniques (ACM SIGGRAPH '07)*, vol. 51, August 2007.

[15] M. I. Smith and J. P. Heather, "Review of image fusion technology," in *Proceedings of the Defense and Security Symposium*, vol. 5782, pp. 29–45, Orlando, Fla, USA, 2005.

[16] G. W. Larson, H. Rushmeier, and C. Piatko, "A visibility matching tone reproduction operator for high dynamic range scenes," *IEEE Transactions on Visualization and Computer Graphics*, vol. 3, no. 4, pp. 291–306, 1997.

[17] F. Drago, K. Myszkowski, T. Annen, and N. Chiba, "Adaptive logarithmic mapping for displaying high contrast scenes," *Computer Graphics Forum*, vol. 22, no. 3, pp. 419–426, 2003.

[18] E. Reinhard and K. Devlin, "Dynamic range reduction inspired by photoreceptor physiology," *IEEE Transactions on Visualization and Computer Graphics*, vol. 11, no. 1, pp. 13–24, 2005.

[19] Y. Li, L. Sharan, and E. H. Adelson, "Compressing and Companding high dynamic range images with subband architectures," *ACM Transactions on Graphics*, vol. 24, no. 3, pp. 836–844, 2005.

[20] R. Fattal, D. Lischinski, and M. Werman, "Gradient domain high dynamic range compression," *ACM Transactions on Graphics*, vol. 21, no. 3, pp. 249–256, 2002.

[21] F. Durand and J. Dorsey, "Fast bilateral Filtering for the display of high dynamic range images," *ACM Transaction on Graphics*, vol. 21, no. 3, pp. 257–266, 2002.

[22] W. F. Lee, T. Y. Lin, M. L. Chu, T. H. Huang, and H. H. Chen, "Perception-based high dynamic range compression in gradient domain," in *Proceedings of the IEEE International Conference on Image Processing (ICIP '09)*, pp. 1805–1808, November 2009.

[23] J. M. Ogden, E. H. Adelson, J. R. Bergen, and P. J. Burt, "Pyramid based computer graphics," *RCA Engineer*, vol. 30, no. 5, pp. 4–15, 1985.

[24] A. Agrawal, R. Raskar, S. K. Nayar, and Y. Li, "Removing photography artifacts using gradient projection and flash exposure

sampling," *ACM Transaction on Graphics*, vol. 24, no. 3, pp. 828–835, 2005.

[25] G. Petschnigg, R. Szeliski, M. Agrawala, M. F. Cohen, H. Hoppe, and K. Toyama, "Digital photography with flash and no-flash image pairs," *ACM Transaction on Graphics*, vol. 23, no. 3, pp. 664–672, 2004.

[26] S. Li and B. Yang, "Multifocus image fusion using region segmentation and spatial frequency," *Image and Vision Computing*, vol. 26, no. 7, pp. 971–979, 2008.

[27] J. H. Adu and M. Wang, "Multi-focus image fusion based on WNMF and focal point analysis," *Journal of Convergence Information Technology*, vol. 6, no. 7, pp. 109–117, 2011.

[28] S. Raman and S. Chaudhuri, "Bilateral filter based compositing for variable exposure photography," in *Proceedings of Eurographics*, Munich, Germany, 2009.

[29] A. Goshtasby, "Fusion of multi-exposure images," *Image and Vision Computing*, vol. 23, pp. 611–618, 2005.

[30] R. Szeliski, "System and process for improving the uniformity of the exposure and tone of a digital image," U. S. Patent No. 6687400, 2004.

[31] Y. Zhao, J. Shen, and Y. He, "Subband architecture based exposure fusion," in *Proceedings of the 4th Pacific-Rim Symposium on Image and Video Technology (PSIVT '10)*, pp. 501–506, Singapore, November 2010.

[32] S. G. Mallat, "A theory for multiresolution signal decomposition: The wavelet representation," *IEEE Transactions on Pattern Analysis and Machine Intelligence*, vol. 11, no. 4, pp. 674–693, 1989.

[33] A. L. da Cunha, J. Zhou, and M. N. Do, "The nonsubsampled contourlet transform: Theory, design, and applications," *IEEE Transactions on Image Processing*, vol. 15, no. 10, pp. 3089–3101, 2006.

[34] M. N. Do and M. Vetterli, "The contourlet transform: An efficient directional multiresolution image representation," *IEEE Transactions on Image Processing*, vol. 14, no. 12, pp. 2091–2106, 2005.

[35] M. J. Black, G. Sapiro, D. H. Marimont, and D. Heeger, "Robust anisotropic diffusion," *IEEE Transactions on Image Processing*, vol. 7, no. 3, pp. 421–432, 1998.

[36] Z. Farbman, R. Fattal, D. Lischinski, and R. Szeliski, "Edge-preserving decompositions for multi-scale tone and detail manipulation," *ACM Transactions on Graphics*, vol. 27, no. 3, article 67, 2008.

[37] C. Wen, G. Gao, and Z. Chen, "Multiresolution model for image denoising based on total least squares," in *Proceedings of the 4th International Conference on Fuzzy Systems and Knowledge Discovery (FSKD '07)*, pp. 622–626, August 2007.

[38] S. Liu, "Adaptive scalar and vector median filtering of noisy colour images based on noise estimation," *IET Image Processing*, vol. 5, no. 6, pp. 541–553, 2011.

[39] D. N. Vizireanu, S. Halunga, and G. Marghescu, "Morphological skeleton decomposition interframe interpolation method," *Journal of Electronic Imaging*, vol. 19, no. 2, Article ID 023018, pp. 1–3, 2010.

[40] K. He, J. Sun, and X. Tang, "Guided image filtering," in *Proceedings of the ECCV*, vol. 6311 of *Lecture Notes in Computer Science*, pp. 1–14, Springer, 2010.

[41] S. Paris, P. Kornprobst, J. Tumblin, and F. Durand, "Bilateral filtering: theory and applications," *Foundations and Trends in, Computer Graphics and Vision*, vol. 4, no. 1, pp. 1–73, 2008.

[42] C. Tomasi and R. Manduchi, "Bilateral filtering for gray and color images," in *Proceedings of the ICCV*, pp. 839–846, IEEE Computer Society, 1998.

[43] J. Canny, "Finding edges and lines in images," Tech. Rep. 720, MIT, Artificial Intelligence Laboratory, 1983.

[44] J. Shen, Y. Zhao, and Y. He, "Detail-preserving exposure fusion using subband architecture," *The Visual Computer*, vol. 28, no. 5, pp. 463–473, 2012.

[45] A. A. Minai and R. D. Williams, "On the derivatives of the sigmoid," *Neural Networks*, vol. 6, no. 6, pp. 845–853, 1993.

[46] R. Shen, I. Cheng, J. Shi, and A. Basu, "Generalized random walks for fusion of multi-exposure images," *IEEE Transactions on Image Processing*, vol. 99, pp. 3634–3646, 2011.

[47] W. Zhang and W.-K. Cham, "Gradient-directed multiexposure composition," *IEEE Transaction on Image Processing*, vol. 21, no. 4, pp. 2318–2323, 2012.

[48] J. Tumblin and G. Turk, "LCIS: a boundary hierarchy for detail-preserving contrast reduction," in *Proceedings of the ACM SIGGRAPH*, pp. 83–90, A. Rockwood, 1999.

[49] P. Hodáková, I. Perfilieva, M. Dǎnková, and M. Vajgl, "Ftransform based image fusion," in *Image Fusion*, O. Ukimura, Ed., pp. 3–22, InTech, Rijeka, Croatia, 2011.

[50] Z. Wang, H. R. Sheikh, and A. C. Bovik, "No reference perceptual quality assessment of JPEG compressed images," in *Proceedings of the International Conference on Image Processing (ICIP '02)*, pp. 477–480, September 2002.